\documentclass[aps,prl,twocolumn,groupedaddress,showkeys,showpacs,superscriptaddress,floatfix]{revtex4-2}
\usepackage{amsmath}
\usepackage{amssymb}
\usepackage{graphicx}
\usepackage{color}

\begin{document}

%\preprint{}

\title{ac-locking of thermally-induced sine-Gordon breathers}


%\author{}
%\thanks{}
%\affiliation{}

\author{Duilio De Santis}
\email[]{duilio.desantis@unipa.it}
\affiliation{Dipartimento di Fisica e Chimica ``E.~Segr\`{e}", Group of Interdisciplinary Theoretical Physics, Università degli Studi di Palermo, I-90128 Palermo, Italy}

\author{Claudio Guarcello} 
%\email[]{cguarcello@unisa.it}
\affiliation{Dipartimento di Fisica ``E.~R.~Caianiello", Università degli Studi di Salerno, I-84084 Fisciano, Salerno, Italy}
\affiliation{INFN, Sezione di Napoli, Gruppo Collegato di Salerno - Complesso Universitario di Monte S. Angelo, I-80126 Napoli, Italy}

\author{Bernardo Spagnolo}
%\email[]{bernardo.spagnolo@unipa.it}
\affiliation{Dipartimento di Fisica e Chimica ``E.~Segr\`{e}", Group of Interdisciplinary Theoretical Physics, Università degli Studi di Palermo, I-90128 Palermo, Italy}
\affiliation{Radiophysics Department, Lobachevsky State University, 603950 Nizhniy Novgorod, Russia}

\author{Angelo Carollo}
%\email[]{angelo.carollo@unipa.it}
\affiliation{Dipartimento di Fisica e Chimica ``E.~Segr\`{e}", Group of Interdisciplinary Theoretical Physics, Università degli Studi di Palermo, I-90128 Palermo, Italy}

\author{Davide Valenti}
%\email[]{davide.valenti@unipa.it}
\affiliation{Dipartimento di Fisica e Chimica ``E.~Segr\`{e}", Group of Interdisciplinary Theoretical Physics, Università degli Studi di Palermo, I-90128 Palermo, Italy}


\date{\today}

\begin{abstract}

A complete framework for exciting and detecting thermally-induced, stabilized sine-Gordon breathers in ac-driven long Josephson junctions is developed. The formation of long-time stable breathers locked to the ac source occurs for a sufficiently high temperature. The latter emerges as a powerful control parameter, allowing for the remarkably stable localized modes to appear. Nonmonotonic behaviors of both the breather generation probability and the energy spatial correlations versus the thermal noise strength are found. The junction's resistive switching characteristics provides a clear experimental signature of the breather.

\end{abstract}

%\keywords{}

\maketitle


\emph{Introduction.}{\textemdash}Owing to its simplicity and nonlinear nature, the sine-Gordon~(SG) equation~\cite{Scott_2006} is universally recognized as a fundamental modelling tool within the scientific community~\cite{Jesus_2014}. The SG framework, in fact, provides a very accurate and intuitive viewpoint for a large variety of phenomena occurring in, e.g., gravity and black holes~\cite{Jesus_2014, Villari_2018}, tectonic stress transfer~\cite{Bykov_2014}, biology~\cite{Ivancevic_2013}, superconductivity and Josephson junctions~(JJs)~\cite{Jesus_2014, Tafuri_2019}, Bose-Einstein condensates~\cite{Su_2015}.

A key feature of the SG equation is its rich spectrum of solutions, which includes both kink-type and breather-type solitons~\cite{Scott_2006}. The first are topological excitations which can be visualized as ${ 2 \pi }$-twists in a mechanical chain of linearly coupled pendula~\cite{Scott_2003, Dauxois_2006}. A breather is a space-localized, time-periodic bound state stemming from the kink-antikink attraction~\cite{Scott_2003, Dauxois_2006}.

The long Josephson junction~(LJJ) is a (quasi) one-dimensional, superconductor-based system whose electrodynamics is reliably described by the SG model~\cite{Scott_2006}. Being the subject of many seminal experiments~\cite{Barone_1982, Ustinov_1992, Ustinov_1998} and striking applications~\cite{Ustinov_1998, Soloviev_2015, Guarcello_2017, Guarcello_2018, Wustmann_2020}, this device has played an outstanding role in the spreading of the soliton concept throughout natural and applied sciences~\cite{Scott_2003, Dauxois_2006, Jesus_2014}. In LJJs, a kink represents a magnetic flux quantum ${ \Phi_0 }$~\cite{Scott_2006}, induced by a supercurrent loop, whose properties reflect into the \emph{I}-\emph{V} characteristic of the junction~\cite{Barone_1982, Ustinov_1992, Ustinov_1998}.

Due to its nontopological structure, mastering the breather's physics is a very tough challenge. In particular, experimental evidence of this oscillating state has yet to be provided in LJJs, despite the numerous investigations on the matter~\cite{Lomdahl_1984, Kivshar_1989, Jensen_1992, Gulevich_2012, De_Santis_2022, De_Santis_2022_CNSNS, De_Santis_2022_NES_arxiv}, primarily due to its friction-triggered radiative decay and its elusiveness with respect to \emph{I}-\emph{V} measurements~\cite{Gulevich_2012, Monaco_2019}. The Josephson breather's detection would, therefore, solve a long-lasting problem in nonlinear science, but it would also pave the way for several applications in, e.g., information transmission~\cite{Macias-Diaz_2007}, quantum computation~\cite{Fujii_2008}, generation of THz radiation~\cite{Krasnov_2011}.

Previous works (e.g., see Ref.~\cite{Jensen_1992}), analyzed the stabilization of stationary SG breathers via ac-driving, with specific ad-hoc initial conditions. Such a scenario, however, has remained experimentally unexplored. This is presumably due to the practical difficulties in creating persistent breather states, given the stabilization effect's crucial dependence on the initial condition. Moreover, the phenomenon's robustness against thermal fluctuations has not been addressed so far. 

On the other hand, the little discussed topic of breathers in a noisy environment has recently gained attention~\cite{Bodo_2009, Calini_2014, De_Santis_2022, De_Santis_2022_CNSNS, De_Santis_2022_NES_arxiv}, and positive stochastically-induced effects on both the generation and the dynamics of these nonlinear waves have been demonstrated. The present manuscript thus examines a lossy, ac-driven LJJ in the presence of thermal noise. The emergence of long-time stable breathers locked to the sinusoidal force is observed for a sufficiently high temperature. The latter is, consequently, a powerful control parameter, allowing for the localized modes to appear, while not endangering their persistence. The achievement of both the creation and the stabilization in a single effort should not be overlooked, given the multistability of the SG system, responsible for the possible emergence of kink-antikink pairs.

As a result, both the probability of exciting solely breathers and the energy spatial correlations are seen to behave nonmonotonically versus the noise strength. Furthermore, at fixed noise intensity, the excitation probability is evaluated in the ac frequency-amplitude space, illustrating the reliability of the approach for different breathing frequencies. A much-awaited, clear experimental signature of the stabilized bound state is finally found in the junction's resistive switching characteristics.

Note that, although the Josephson realm provides a solid physical background for this letter, the formalism is quite general, and an interdisciplinary flavor characterizes the analysis. In other words, since many complex and apparently different phenomena~\cite{Jesus_2014, Villari_2018, Bykov_2014, Ivancevic_2013, Tafuri_2019, Su_2015} can be understood through the lens of the SG model, significant insights into its fundamental excitations have a wide scope within the scientific community. The topic of SG breathers is indeed of general interest: from DNA systems~\cite{Liu_2021} and structural geology~\cite{Zalohar_2020} to high-${ T_c }$ superconductivity~\cite{Dienst_2013}.

Other examples of breather-type states intensely studied are: polygonal breathers~\cite{Alperin_2022}, matter–wave breathers~\cite{Luo_2020}, breather wave molecules~\cite{Xu_2019}, rotobreathers in JJ ladders~\cite{Trias_2000, Binder_2000}. Besides, in JJ parallel arrays, the theoretically-predicted oscillobreathers, due to their rapid pulsations, have eluded an experimental verification for decades~\cite{Mazo_2003}. Exploring the noisy, ac-driven scenario in a fashion similar to that presented here could lead to interesting developments even in the discrete world~\cite{Hennig_2008, Cubero_2009}.

\emph{The model.}{\textemdash}Taking into account dissipation, an ac current uniformly distributed in space, and thermal fluctuations, the equation of motion for the LJJ reads~\cite{Barone_1982, Castellano_1996}
\begin{equation}
\label{eqn:1}
\varphi_{xx} - \varphi_{tt} - \alpha \varphi_{t} = \sin \varphi - \eta \sin (\omega t) - \gamma_T (x, t) ,
\end{equation}
with ${ \varphi (x, t) }$ indicating the phase difference between the two superconducting wave functions (the notation ${ \partial \varphi / \partial x = \varphi_x }$ is used throughout). The friction coefficient ${ \alpha = G / \left( \omega_p C \right) }$ is defined in terms of the effective normal conductance $ G $, the capacitance per unit length ${ C }$, and the Josephson plasma frequency ${ \omega_p = \sqrt{ 2 \pi J_c / \left( \Phi_0 C \right) } }$, with respect to which frequency is normalized in Eq.~\eqref{eqn:1} ($ J_c $ is the critical Josephson current density)~\cite{Barone_1982}. The spatial length scale is the Josephson penetration depth ${ \lambda_J = \sqrt{ \Phi_0 / \left( 2 \pi J_c L_P \right)} }$, where ${ L_P }$ is the inductance per unit length. Moreover, ${ \omega }$ and ${ \eta }$ are, respectively, the normalized frequency and amplitude of the external ac driving (${ \eta }$ is given in units of $ J_c $), and ${ \gamma_{T} (x, t) }$ is a Gaussian, zero-average noise source with the correlation function
\begin{equation}
\label{eqn:2}
\langle \gamma_{T}(x_1, t_1) \gamma_{T}(x_2, t_2) \rangle = 2 \alpha \Gamma \delta (x_1 - x_2) \delta (t_1 - t_2) ,
\end{equation}
in which ${ \Gamma = 2 e k_B T / \left( \hbar J_c \lambda_J \right) }$ is the noise strength, proportional to the absolute temperature $ T $, $ e $ is the electron charge, $ k_B $ is the Boltzmann constant, and $ \hbar $ is the reduced Planck constant. Equation~\eqref{eqn:1} is numerically integrated via an implicit finite-difference scheme, in the spatio-temporal domain ${ [-l/2, l/2] \times [0, \mathcal{T}] }$, with initial conditions
\begin{equation}
\label{eqn:3}
\varphi (x, 0) = \varphi_t (x, 0) = 0 ,
\end{equation}
and periodic boundary conditions
\begin{equation}
\label{eqn:4}
\varphi (-l/2, t) = \varphi (l/2, t) ,
\end{equation}
the latter corresponding to an annular-geometry LJJ~\cite{Ustinov_1992}. More details, including the approximation of the stochastic term, can be found in~\footnote[1]{See Supplemental Material for more information on the SG equation, the numerical techniques, the energy-based analysis of the spatial correlations, and a discussion of the typical timescale of the generation events.}. In what follows, the junction length is ${ l = 50 }$, the damping parameter is ${ \alpha = 0.2 }$~\cite{Guarcello_2017}, and ${ \omega < 1 }$, since below-plasma frequencies are those natural to SG breathers~\cite{Scott_2003, Dauxois_2006}.

\begin{figure}[t!!]
\includegraphics[width=\columnwidth]{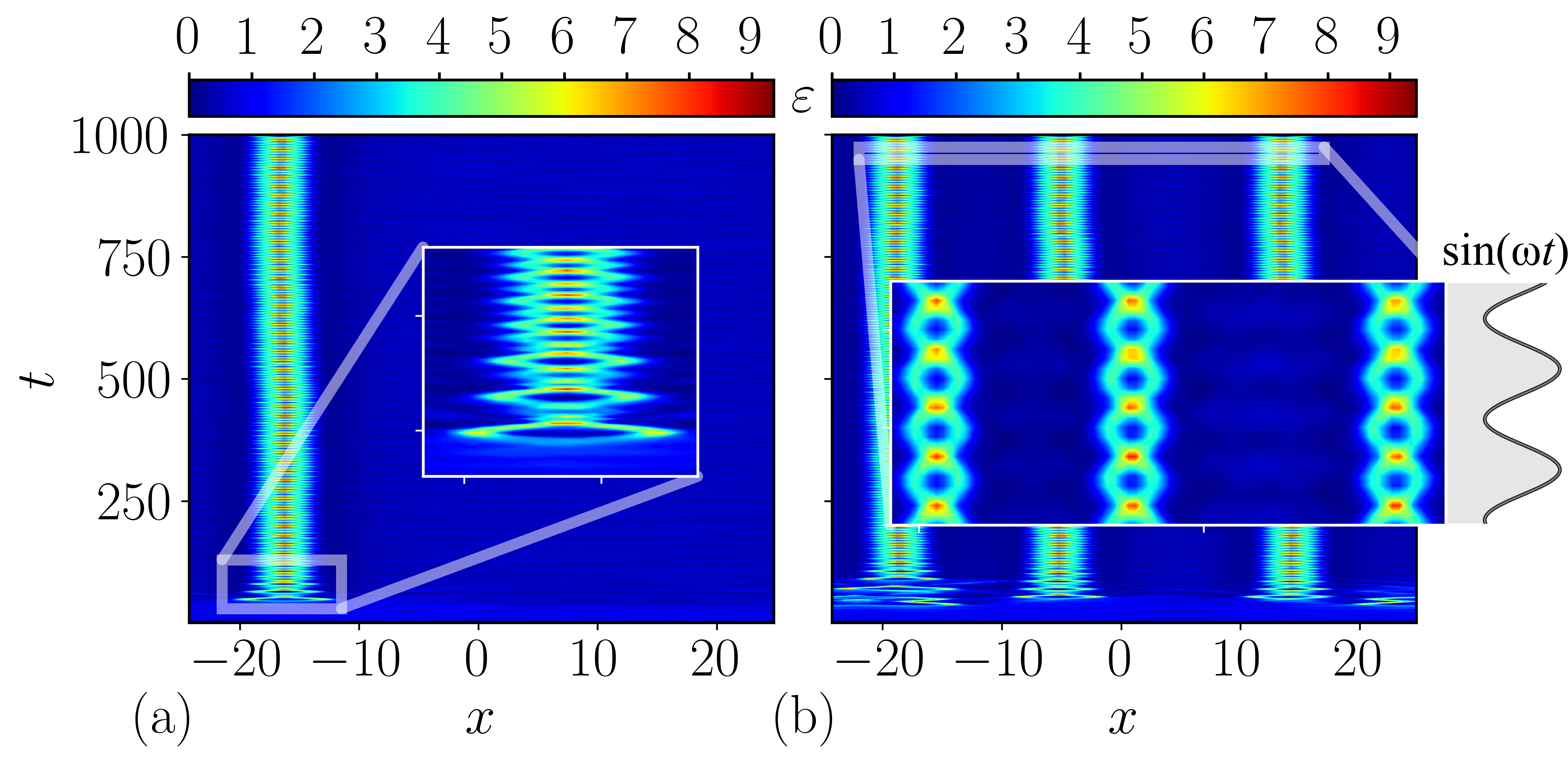}
\caption{Two simulated energy density profiles ${ \varepsilon (x, t) = (\varphi_t^2 + \varphi_x^2)/2 + 1 - \cos \varphi }$~\cite{Scott_2003, Dauxois_2006}. In panel~(a), the spatio-temporal region ${ [-21.5, -11.5] \times [30, 130] }$ is magnified to better appreciate both the formation and the first few oscillations of a single breather located at ${ x \approx -16.5 }$. In panel~(b), the inset focuses on ${ [-22, 17] \times [950, 975] }$ to illustrate the ac-locking of multiple nonlinear modes. Parameter values: ${ \mathcal{T} = 1000 }$ (observation time), ${ \omega = 0.6 }$, ${ \eta = 0.59 }$, and ${ \Gamma = 5 \times 10^{-4} }$.}
\label{fig:1}
\end{figure}
\emph{Noise-induced, stabilized breathers.}{\textemdash}Figure~\ref{fig:1} displays two simulated energy density profiles ${ \varepsilon (x, t) = (\varphi_t^2 + \varphi_x^2)/2 + 1 - \cos \varphi }$~\cite{Scott_2003, Dauxois_2006}. Both panels demonstrate that, in the presence of thermal fluctutations and ac forcing, remarkably stable breather excitations can form in the junction. In a purely dissipative case, breathers radiatively decay within ${ \sim 1 / \alpha = 5 }$~\cite{De_Santis_2022_NES_arxiv}, a lifetime which is surpassed by multiple orders of magnitude here. Note also the stability of the modes with respect to the position, i.e., their centers do not drift away from the originary positions [${ x \approx -16.5 }$ in Fig.~\ref{fig:1}(a)] over hundreds of oscillations, despite the noise influence. These interesting features hold widely among the different realizations. One or more breathers typically appear in random spots within a few driving cycles (${ t \approx 50 }$ in Fig.~\ref{fig:1}). After a transient, a state similar to that of Fig.~\ref{fig:1}, stable over very long times~\footnote[2]{The choice ${ \mathcal{T} = 1000 }$ in Fig.~\ref{fig:1} was made for visualization purposes. No radiative decay was observed even for higher ${ \mathcal{T} }$ values.}, eventually sets in.

Further information regarding the stabilized oscillatory modes is perhaps useful here: (i)~their breathing cycles are locked to the external ac force [Fig.~\ref{fig:1}(b), inset]; (ii)~they are strongly localized in space, over the characteristic length ${ \lambda_b \left( \omega \right) = 1 / \sqrt{1 - \omega^2} }$~\cite{Scott_2003, Dauxois_2006}, i.e., the width of an unperturbed breather at frequency ${ \omega_b = \omega }$; (iii)~their amplitude is ${ \gtrsim A_b \left( \omega \right) = 4 \arctan \left( \sqrt{1 - \omega^2} / \omega \right) }$~\cite{Scott_2003, Dauxois_2006}, which is that of an unperturbed breather at the driving's frequency ${ \omega }$~\footnote[3]{Furthermore, a test was run at ${ \Gamma = 0 }$, starting from an exact breather at frequency ${ \omega }$, in the presence of the two perturbations ${ \alpha \varphi_{t} }$ and ${ \eta \sin (\omega t) }$. The breather was seen to adjust its amplitude to that observed for the same ${ \alpha }$, ${ \omega }$, and ${ \eta }$ values, in the case of noise-induced formation.}.

\begin{figure}[t!!]
\includegraphics[width=\columnwidth]{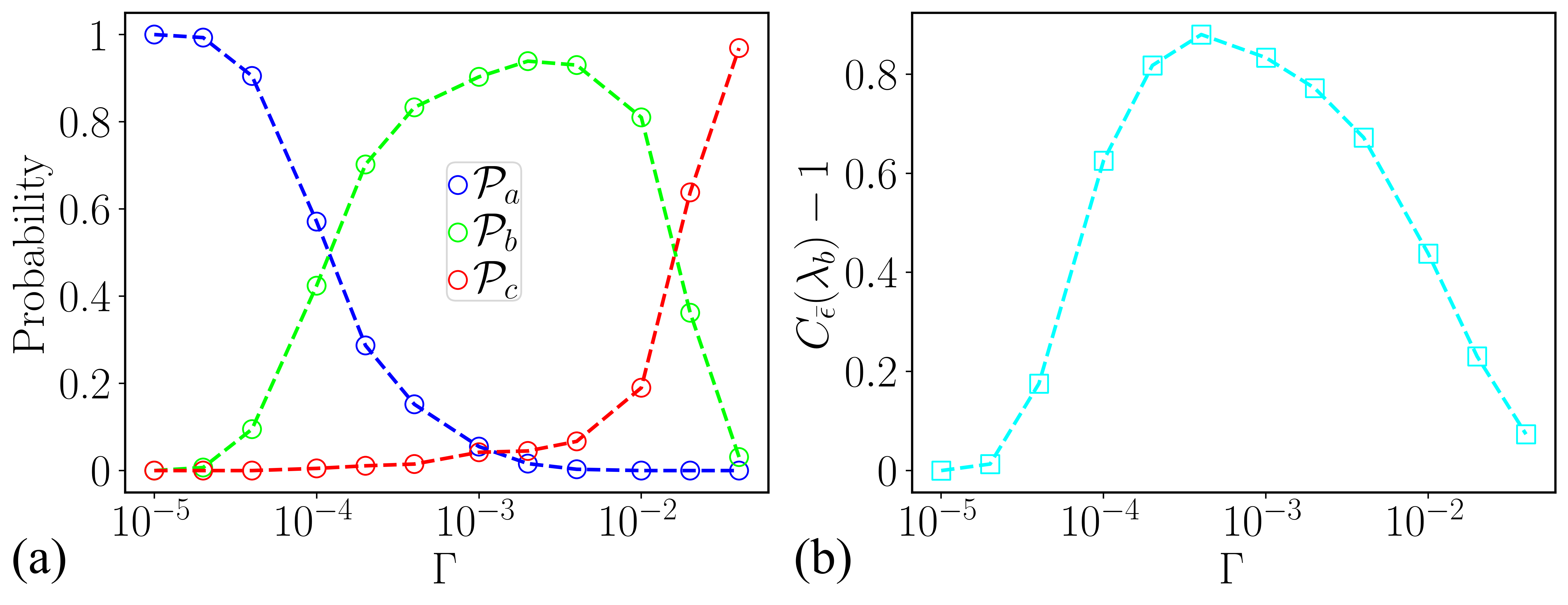}
\caption{(a): Probability of having no excitations (${ \mathcal{P}_a }$, blue), breathers only (${ \mathcal{P}_b }$, green), and at least a free kink-antikink couple (${ \mathcal{P}_c }$, red) versus ${ \Gamma }$. (b): Energy-based coefficient of spatial correlation, see Eq.~\eqref{eqn:S9}, as a function of ${ \Gamma }$. Parameter values: ${ \mathcal{T} = 500 }$, ${ \omega = 0.6 }$, ${ \eta = 0.59 }$, and ${ N = 1000 }$.}
\label{fig:2}
\end{figure}
Keeping the parameter values ${ \omega = 0.6 }$ and ${ \eta = 0.59 }$ as in Fig.~\ref{fig:1}, the junction's response versus the noise strength ${ \Gamma \in [10^{-5}, 4 \times 10^{-2}] }$~\cite{Guarcello_2017} is now explored, for ${ \mathcal{T} = 500 }$ and ${ N = 1000 }$ realizations. Specifically, simulating for a time long enough to let the generation events to unravel, the final state of each run is classified as follows: (a)~no excitations, if the phase profile is essentially flat over the spatial domain; (b)~breathers only, if the observed modes' amplitudes lie between ${ A_b }$ and ${ 2 \pi }$, the latter being the phase value associated with kink-type structures~\cite{Scott_2003, Dauxois_2006}; (c)~at least a free kink-antikink couple, if at least a ${ 2 \pi }$-step excitation is present.

As illustrated in Fig.~\ref{fig:2}(a), for the lower ${ \Gamma }$ values, the probability ${ \mathcal{P}_a }$ of having no excitations is ${ 1 }$ (see the blue circles). As the noise intensity is increased, a new scenario soars, that of breather-only formation. Indeed, for ${ \Gamma }$ roughly in ${ [5 \times 10^{-4}, 10^{-2}] }$, the corresponding probability ${ \mathcal{P}_b }$ is ${ \gtrsim 0.9 }$ (see the green circles). This provides a rather wide range of working temperatures for the current approach. The stochastic influence eventually becomes disruptive for the oscillatory bound state, and the kink-antikink regime takes over for ${ \Gamma > 10^{-2} }$ (see the red circles, ${ \mathcal{P}_c }$). The probability of exciting solely breathers therefore exhibits a nonmonotonicity versus ${ \Gamma }$, highlighting the crucial role of the temperature as a control parameter in the setup. In this regard, the fact that thermal noise can allow for the formation process, without compromising the long-time stability of the breathers, is noteworthy.

Furthermore, the energy spatial correlation evaluated at the characteristic scale ${ \lambda_b }$~\footnote[1]{}
\begin{equation}
\label{eqn:S9}
C_{\bar{\varepsilon}} (\lambda_b) \propto \frac{\left\langle \int \bar{\varepsilon}(x) \bar{\varepsilon}(x + \lambda_b) dx \right\rangle}{{\left\langle \int \bar{\varepsilon}(x) dx \right\rangle}^2} ,
\end{equation}
where ${ \bar{\varepsilon}(x) }$ is the time-averaged energy density, shows a nonmonotonic behavior as a function of ${ \Gamma }$ [see Fig.~\ref{fig:2}(b)]. Thus, an appropriate amount of environmental noise, instead of degradation, enhances the junction's sensitivity to the external force, leading to nontrivial spatial correlations{\textemdash}a somewhat counter-intuitive outcome. The noise amplitude also impacts the typical timescale of the generation events: for stronger fluctuations, they occur earlier in the simulations. This aspect is quantitatively addressed in~\footnote[1]{}.

\begin{figure}[t!!]
\includegraphics[width=0.75\columnwidth]{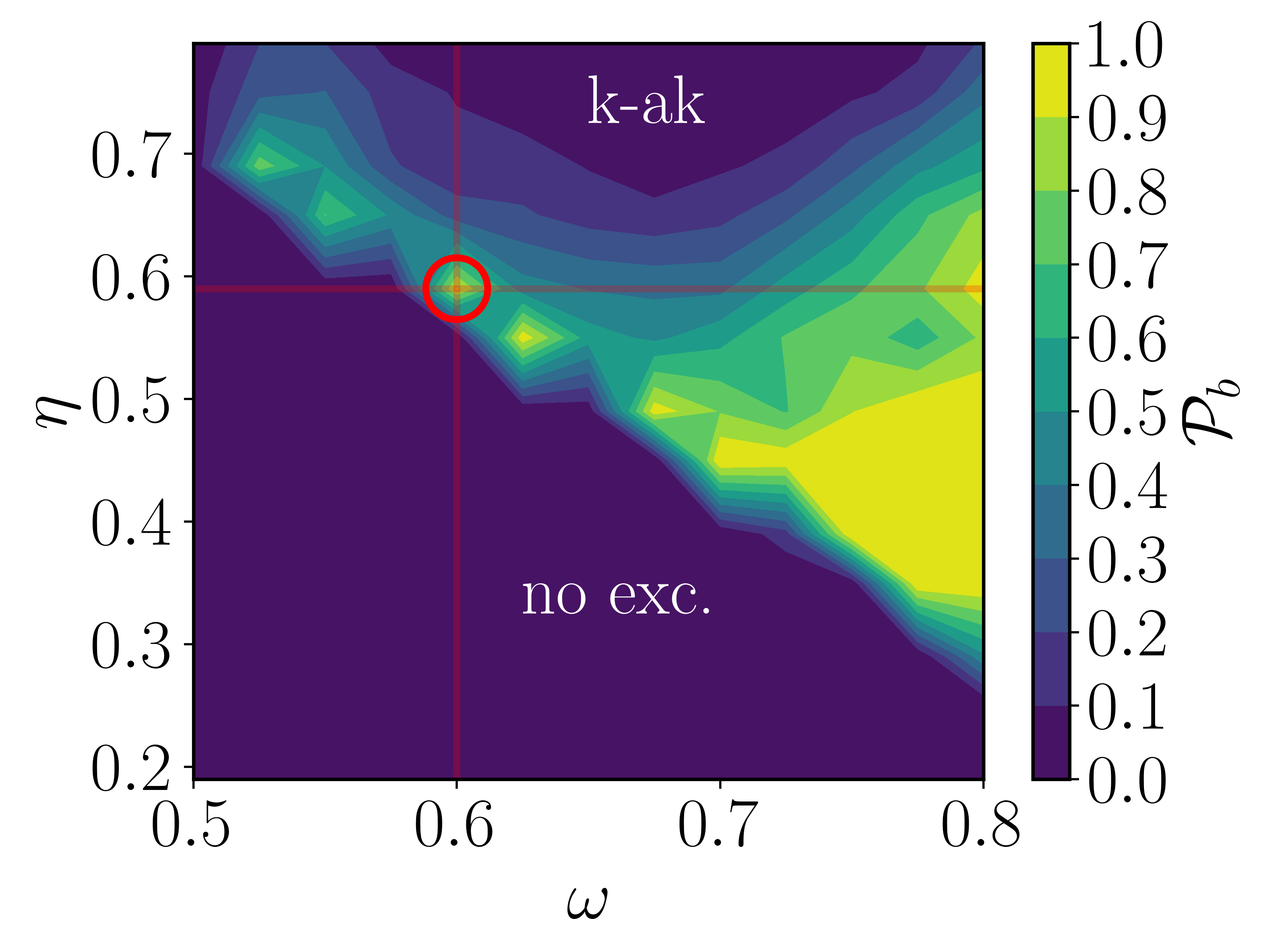}
\caption{Probability of generating solely breathers in the ${ (\omega, \eta) }$ parameter space. The red circle identifies the combination ${ \omega = 0.6 }$ and ${ \eta = 0.59 }$. Parameter values: ${ \mathcal{T} = 500 }$, ${ \Gamma = 5 \times 10^{-3} }$, and ${ N = 500 }$.}
\label{fig:3}
\end{figure}
It is now important to examine, at fixed ${ \Gamma > 0 }$, the behavior of the breather-only generation probability ${ \mathcal{P}_b }$ in the frequency-amplitude parameter space~\footnote[4]{In the absence of thermal noise, no formation of nonlinear modes occurs, regardless of the ${ \omega }$ and ${ \eta }$ values.}. To cope with such a heavy computational task, ${ N = 500 }$ runs are performed for each ${ (\omega, \eta) }$ pair, focusing on ${ \omega \in [0.5, 0.8] }$ and ${ \eta \in [0.2, 0.8] }$, with ${ \Delta \omega = 0.02 }$ and ${ \Delta \eta = 0.05 }$. The simulation time and noise amplitude are ${ \mathcal{T} = 500 }$ and ${ \Gamma = 5 \times 10^{-3} }$, respectively.

Figure~\ref{fig:3} shows that several high-${ \mathcal{P}_b \left( \omega, \eta \right) }$ (green, yellow) areas exist for breather-only formation. Note that, for the scenario of Fig.~\ref{fig:1} to occur, the combined action of noise and the deterministic force must provide an energy of the order of ${ E_b \left( \omega \right) = 16 \sqrt{ 1 - \omega^2 } }$~\cite{Scott_2003, Dauxois_2006}, i.e., that expected for a breather at frequency ${ \omega }$, without breaking up any of the subsequent kink-antikink bonds. Two reasons are behind the low-probability (purple) region. The first one, for ${ \eta \gtrsim 0.7 }$ (see Fig.~\ref{fig:3}), is the kink-antikink (k-ak) regime, associated to an excess of energy input. For the remaining purple ${ (\omega, \eta) }$ area, no excitations are observed. One may notice that, at lower ${ \omega }$ values, higher amplitudes ${ \eta }$ are needed to excite the nonlinear breathing states. This is qualitatively explained by the above expression of ${ E_b \left( \omega \right) }$, which implies that breathers with lower frequencies require more energy.

Another topic worth discussing is the system's topology and its influence on the examined phenomenon. Due to Eq.~\eqref{eqn:4}, the (initially null) topological charge is conserved, thus no unpaired kinks/antikinks can arise. By contrast, for Neumann-type boundary conditions, i.e., for an overlap-geometry LJJ~\cite{De_Santis_2022, De_Santis_2022_CNSNS}, single kinks/antikinks can emerge at the borders, usually forming bound states with their virtual counterparts~\cite{Costabile_1978, Jensen_1992}{\textemdash}what one may call edge-breathers. The latter case was extensively analyzed as well (not shown here), and the overall picture is not drastically altered. The difference is that in the periodic framework, i.e., annular LJJs, there are no preferred locations for the emergence of breather states, whereas in the Neumann case, i.e., overlap LJJs, edge-breathers, being essentially single-soliton modes, are more likely observed since they provide an energetic advantage.

\emph{Detection.}{\textemdash}The lowest dc current value to break up an unperturbed breather into a kink-antikink pair crucially depends on its phase~\cite{Lomdahl_1984, Gulevich_2012}. Starting from this insight, and taking full advantage of the developed setup, a much-awaited, clear experimental signature of the oscillatory bound state is provided.

The parameters ${ \omega = 0.6 }$, ${ \eta = 0.59 }$, and ${ \Gamma = 5 \times 10^{-4} }$ are selected here to work with a highly favorable breather formation scenario (see Figs.~\ref{fig:2}~and~\ref{fig:3}). The physical idea behind the detection scheme is quite simple: (i)~excite stabilized breathers; (ii)~embed their properties into the switching characteristics of the device by destroying them at different stages of their oscillation cycle. More precisely, the ac-driven LJJ is first let to evolve up to ${ t = \left( t^\star + \tau \right) }$, where ${ t^\star }$ is a time much greater than that typical for the occurrence of the generation events, and ${ \tau }$ is an arbitrary (time) displacement. With the chosen values of ${ \omega }$, ${ \eta }$, and ${ \Gamma }$, breathers emerge roughly within ${ t = 50 }$ (see Fig.~\ref{fig:1} and ~\footnote[1]{}), thus ${ t^\star = 250 }$ is taken to allow the system to reach its long-time stable configuration. Next, the smooth current bias ${ \gamma \left\lbrace 1 - \exp[ - 0.1 ( t - t^\star - \tau ) ] \right\rbrace }$~\cite{Gulevich_2012} is applied for ${ t > \left( t^\star + \tau \right) }$, while the ac force is slowly turned off, and one should record whether the junction switches to a resistive state{\textemdash}namely, whether the kink-antikink splitting is triggered and a measurable voltage drop appears. The previous steps have then to be repeated a number of times to obtain, for each different ${ \tau }$ value, the minimal current ${ \gamma_{\rm{sw}} }$ leading to a significant switching probability over ${ N }$ realizations, say, ${ P_{\rm{sw}} \geq 0.75 }$.

A few relevant points underlying the above approach should be mentioned. Past proposals with a similar goal~\cite{Gulevich_2012} have encountered the serious issue of dissipation. The modes' stability for ${ t \leq \left( t^\star + \tau \right) }$ practically solves the problem here. Second, as previously mentioned, the breather oscillations are locked to the ac-drive, ensuring that breathers from all the repetitions at fixed ${ \tau }$ arrive in phase at ${ t = \left( t^\star + \tau \right) }$. This is crucial, since the whole idea revolves around breaking up the solitonic bound states at different stages of their oscillation cycle~\footnote[5]{Each `stage' corresponds to a displacement ${ \tau }$, and it has to be replicated multiple times to evaluate ${ P_{\rm{sw}} }$.}. Note also that the randomness in the number of breathers emerging in each realization does not harm the described sequence in any way. Lastly, the slow switch-off of the ac driving for ${ t > \left( t^\star + \tau \right) }$ avoids the simultaneous action of noise, the smooth current bias, and the ac source at full strength. The latter situation, in fact, can potentially lead to additional kink-antikink states that would pretty much take over the switching dynamics of the junction.

\begin{figure}[t!!]
\includegraphics[width=0.65\columnwidth]{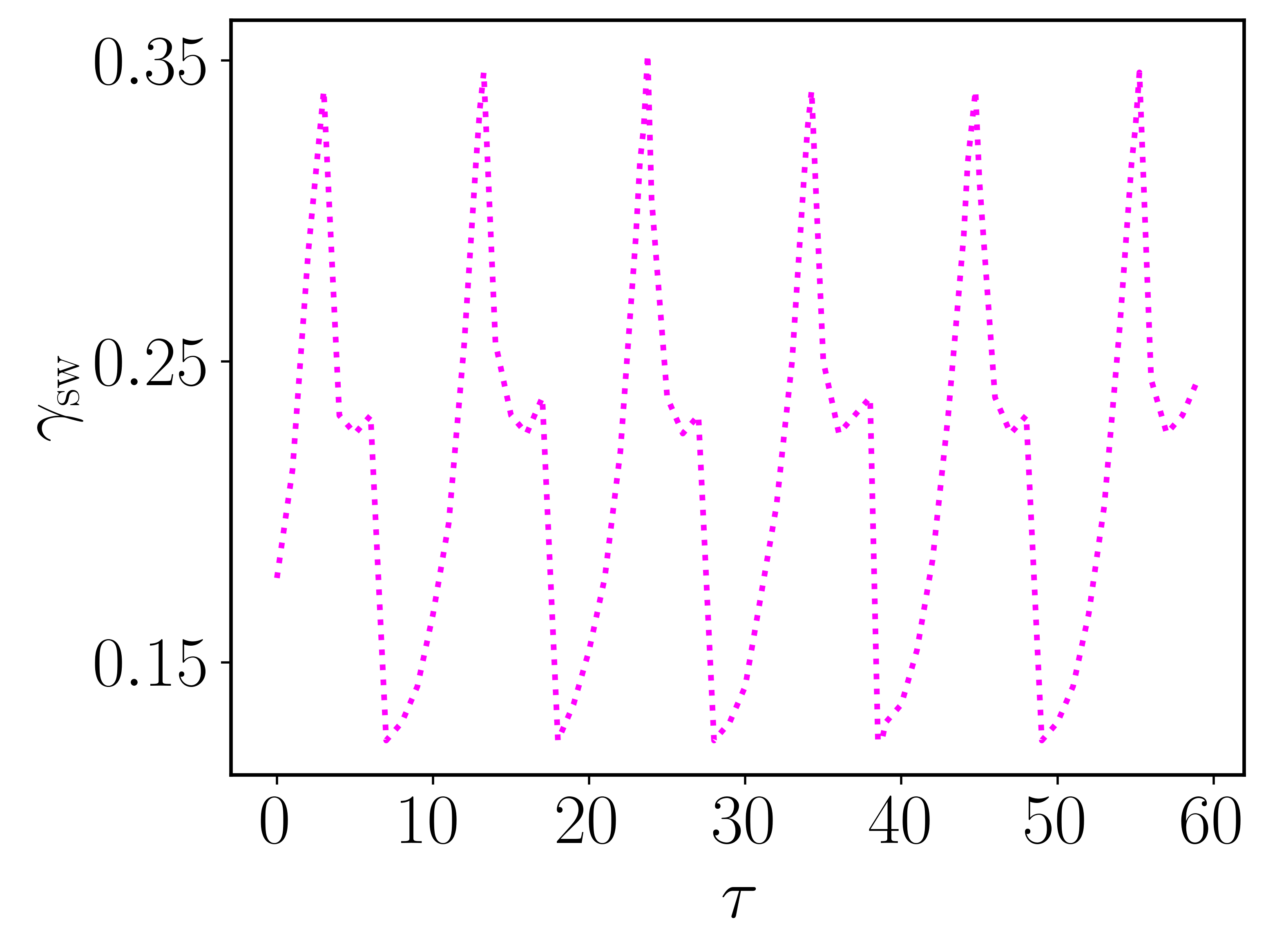}
\caption{Lowest current value ${ \gamma_{\rm{sw}} }$ at which the resistive state is triggered with probability ${ P_{\rm{sw}} \geq 0.75 }$ as a function of the time displacement ${ \tau \in [0, 59] }$. The ac driving's slow switch-off consists in the time-dependent amplitude ${ \eta \exp[ - 0.01 ( t - t^\star - \tau ) ] }$ for ${ t > \left( t^\star + \tau \right) }$. Parameter values: ${ \mathcal{T} = 500 }$, ${ \omega = 0.6 }$, ${ \eta = 0.59 }$, ${ \Gamma = 5 \times 10^{-4} }$, ${ t^\star = 250 }$, and ${ N = 500 }$.}
\label{fig:4}
\end{figure}
The quantity ${ \gamma_{\rm{sw}} (\tau) }$ displays a peculiar oscillatory behavior (see Fig.~\ref{fig:4}). A period approximately equal to ${ 10 \approx 2 \pi / \omega }$ can be appreciated, which reflects the breathing cycle. This outcome is markedly different from that obtained both in the absence of excitations and in a kink-antikink regime, where no sensitivity to the displacement ${ \tau }$ is exhibited. Indeed, in the small-noise case ${ \Gamma = 10^{-5} }$, where essentially no excitations appear [${ \mathcal{P}_a \approx 1 }$ in Fig.~\ref{fig:2}(a)], one gets ${ P_{\rm{sw}} \approx 0 }$ for ${ \gamma \in \left[ 0, 0.4 \right] }$, independently of ${ \tau }$. With ${ \Gamma = 4 \times 10^{-2} }$ [${ \mathcal{P}_c \approx 1 }$ in Fig.~\ref{fig:2}(a), i.e., kink-antikink scenario] the minimal current is ${ \gamma_{\rm{sw}} \approx 0.17 \; \forall \tau }$.

\emph{Conclusions.}{\textemdash}This letter addresses the formation of breathers stable over long times, for sufficiently high temperatures, in ac-driven LJJs. Nonmonotonic behaviors of both the probability of generating solely breathers and the energy spatial correlations are obtained as a function of the noise strength, highlighting the latter's critical role as a control parameter. The efficacy of the phenomenon for different breathing frequencies is demonstrated. Lastly, the breather induces peculiar oscillations into the junction's resistive switching characteristics, which is exploitable to experimentally reveal it.

Preliminary simulations indicate that the results are robust even to static disorder due, e.g., to impurities in the device. It may also be interesting to design a setup where preferred locations for the emergence of breathers can be selected. This could be, reasonably, achieved by locally heating the junction or by means of a spatially-modulated ac force~\cite{Jensen_1992}.

\begin{acknowledgments}

The authors are very grateful to Prof. A. Ustinov for suggesting the topic of breathers in Josephson systems and for
stimulating discussions. DDS gladly acknowledges fruitful discussions with Prof. D. Molteni. Most of the numerical runs were performed on CINECA's machine Galileo100 (Projects: IscrC{\textunderscore}NDJB and IscrB{\textunderscore}3DSBM). DDS, CG, BS, AC, DV acknowledge the support of the Italian Ministry of University and Research (MUR). BS also acknowledges the support of the Government of the Russian Federation through Agreement No. 074-02-2018-330 (2).

\end{acknowledgments}

%\bibliography{NSGB_De_Santis_References}


\end{document}

%\preprint{}

\title{Supplemental Material\\ac-locking of thermally-induced sine-Gordon breathers}


%\author{}
%\thanks{}
%\affiliation{}

\author{Duilio De Santis}
\email[]{duilio.desantis@unipa.it}
\affiliation{Dipartimento di Fisica e Chimica ``E.~Segr\`{e}", Group of Interdisciplinary Theoretical Physics, Università degli Studi di Palermo, I-90128 Palermo, Italy}

\author{Claudio Guarcello} 
%\email[]{cguarcello@unisa.it}
\affiliation{Dipartimento di Fisica ``E.~R.~Caianiello", Università degli Studi di Salerno, I-84084 Fisciano, Salerno, Italy}
\affiliation{INFN, Sezione di Napoli, Gruppo Collegato di Salerno - Complesso Universitario di Monte S. Angelo, I-80126 Napoli, Italy}

\author{Bernardo Spagnolo}
%\email[]{bernardo.spagnolo@unipa.it}
\affiliation{Dipartimento di Fisica e Chimica ``E.~Segr\`{e}", Group of Interdisciplinary Theoretical Physics, Università degli Studi di Palermo, I-90128 Palermo, Italy}
\affiliation{Radiophysics Department, Lobachevsky State University, 603950 Nizhniy Novgorod, Russia}

\author{Angelo Carollo}
%\email[]{angelo.carollo@unipa.it}
\affiliation{Dipartimento di Fisica e Chimica ``E.~Segr\`{e}", Group of Interdisciplinary Theoretical Physics, Università degli Studi di Palermo, I-90128 Palermo, Italy}

\author{Davide Valenti}
%\email[]{davide.valenti@unipa.it}
\affiliation{Dipartimento di Fisica e Chimica ``E.~Segr\`{e}", Group of Interdisciplinary Theoretical Physics, Università degli Studi di Palermo, I-90128 Palermo, Italy}


\date{\today}

%\begin{abstract}
%
%\end{abstract}

%\keywords{}

\maketitle


\section{The sine-Gordon equation}

The sine-Gordon~(SG) equation
\begin{equation}
\label{eqn:S1}
\varphi_{xx} - \varphi_{tt} = \sin \varphi ,
\end{equation}
\sloppy which can be derived from the energy density ${
\varepsilon (x, t) = (\varphi_t^2 + \varphi_x^2)/2 + 1 - \cos \varphi }$, admits topological soliton solutions
\begin{equation}
\label{eqn:S2}
\varphi_{\pm} (x, t) = 4 \arctan \left[ \exp \left( \pm \frac{x - vt}{\sqrt{1 - v^2}} \right) \right]
\end{equation} 
known as kinks ($ \varphi_+ $) and antikinks ($ \varphi_- $). In Eq.~\eqref{eqn:S2}, ${ v < 1 }$ is the constant velocity of the travelling wave, whose relativistic behavior is due to the structure of Eq.~\eqref{eqn:S1}. In the realm of long Josephson junctions~(LJJs), the limiting velocity is the propagation speed of electromagnetic signals in the device, commonly known as the Swihart velocity, ${ \bar{c} = \lambda_J \omega_p = 1 / \sqrt{L_P C} }$, with ${ \lambda_J }$ being the Josephson penetration depth, ${ \omega_p }$ the Josephson plasma frequency, ${ L_P }$ the inductance per unit length, and ${ C }$ the capacitance per unit length~\cite{Barone_1982}.

Breathers are space-localized, time-periodic solutions of Eq.~\eqref{eqn:S1} given by
\begin{equation}
\label{eqn:S3}
\varphi_{b} (x, t) = 4 \arctan \left\lbrace \frac{\sqrt{1 - \omega_b^2}}{\omega_b} \frac{ \sin \left[ \frac{\omega_b \left( t - v_e x \right)}{\sqrt{1 - v_e^2}} \right] } { \cosh \left[ \frac{\sqrt{1 - \omega_b^2} \left( x - v_e t \right)}{\sqrt{1 - v_e^2}} \right] } \right\rbrace ,
\end{equation}
where ${ 0 < \omega_b < 1 }$ and ${ v_e < 1 }$ are, respectively, the excitation's proper frequency and the envelope velocity. Notably, Eq.~\eqref{eqn:S3} can be obtained via analytic continuation from the profile describing the kink-antikink collision. If ${ v_e = 0 }${\textemdash}the most relevant case for the present work{\textemdash}besides the duration of each breathing cycle ${ T_b = 2 \pi / \omega_b }$, the parameter $ \omega_b $ yields the energy ${ E_b = 16 \sqrt{ 1 - \omega_b^2 } }$, the amplitude ${ A_b = 4 \arctan \left( \sqrt{1 - \omega_b^2} / \omega_b \right) }$, and the characteristic length ${ \lambda_b = 1 / \sqrt{1 - \omega_b^2} }$ of the nonlinear mode. Therefore, high (low) energy breathers possess low (high) frequencies and high (low) oscillation amplitudes.

For more information concerning the SG equation, even in the presence of perturbation terms, see Refs.~\cite{Scott_2003, Dauxois_2006} and references therein.

\section{Numerical solution of the perturbed sine-Gordon equation}

An implicit finite-difference scheme is employed to integrate the following perturbed SG equation
\begin{equation}
\label{eqn:S4}
\varphi_{xx} - \varphi_{tt} - \alpha \varphi_{t} = \sin \varphi - \eta \sin (\omega t) ,
\end{equation}
where ${ \alpha }$ is the dissipation coefficient and ${ \omega }$ and ${ \eta }$ are, respectively, the frequency and amplitude of the monochromatic force. More specifically, the spatial domain is divided into $ \mathcal{N} $ cells of length ${ \Delta x = h }$ and the temporal domain into $ \mathcal{M} $ intervals of duration ${ \Delta t = k }$. Within this framework, the restriction of ${ \varphi (x, t) }$ is indicated as ${ \varphi_n^m = \varphi (n h, m k) }$, for ${ n = 1, ..., \mathcal{N} }$ and ${ m = 1, ..., \mathcal{M} }$. Then, one replaces the derivatives in Eq.~\eqref{eqn:S4} with~\cite{Ames_1977}
\begin{equation}
\label{eqn:S5}
\begin{aligned}
& \varphi_x = \frac{1}{2 h} \left( \varphi_{n + 1}^{m} - \varphi_{n - 1}^{m} \right) + O (h^2) , \\
& \varphi_t = \frac{1}{2 k} \left( \varphi_{n}^{m + 1} - \varphi_{n}^{m - 1} \right) + O (k^2) , \\
& \varphi_{xx} = \frac{1}{2 h^2} \left( \varphi_{n + 1}^{m + 1} - 2 \varphi_{n}^{m + 1} + \varphi_{n - 1}^{m + 1} + \varphi_{n + 1}^{m - 1} - 2 \varphi_{n}^{m - 1} + \varphi_{n - 1}^{m - 1} \right) + O (h^2 + k^2) , \\
& \varphi_{tt} = \frac{1}{k^2} \left( \varphi_{n}^{m + 1} - 2 \varphi_{n}^{m} + \varphi_{n}^{m - 1} \right) + O (k^2) .
\end{aligned}
\end{equation}
If ${ O(h^2) }$ and ${ O(k^2) }$ terms are neglected, and both the initial and the boundary conditions are imposed, one gets a system of equations, whose resolution determines the new (unknown) values ${ \varphi_{n}^{m + 1} }$, given the previous ones ${ \varphi_{n}^{m} }$ and ${ \varphi_{n}^{m - 1} }$, with ${ n = 1, ..., \mathcal{N} }$. For periodic boundary conditions, the matrix representing the system is cyclic tridiagonal, i.e., it has nonzero elements only on the diagonal, the subdiagonal, the superdiagonal, and in the corners. The solution is therefore found by combining the Sherman-Morrison formula with Thomas' algorithm (the latter being a simplified form of Gaussian elimination)~\cite{Press_1992}. Moreover, Wilkinson's iterative refinement method is used at each step to prevent the accumulation of rounding errors~\cite{Press_1992}.

According to Ref.~\cite{Tuckwell_2016}, the noisy perturbation ${ \gamma_T (x, t) }$, whose statistical properties are
\begin{equation}
\label{eqn:S6}
\langle \gamma_{T}(x, t) \rangle = 0 \text{\phantom{00}and\phantom{00}} \langle \gamma_{T}(x_1, t_1) \gamma_{T}(x_2, t_2) \rangle = 2 \alpha \Gamma \delta (x_1 - x_2) \delta (t_1 - t_2) ,
\end{equation}
can be numerically handled as
\begin{equation}
\label{eqn:S7}
\sqrt{2 \alpha \Gamma} \frac{W_{n}^{m}}{\sqrt{h k}} ,
\end{equation}
in which ${ \Gamma }$ is the noise strength and ${ W_{n}^{m} }$ are independent normal random variables with zero mean and unit variance.

The computational scheme's precision was tested by systematically varying the values of the space and time steps, and by examining the discrepancy with a variety of analytical SG solutions, such as Eqs.~\eqref{eqn:S2}~and~\eqref{eqn:S3}, in the ${ \alpha = \eta = \Gamma = 0 }$ case.

Throughout the work, the chosen discretization steps are ${ h = k = 0.01 }$.

\section{Energy-based analysis of the spatial correlations}

The energy density ${ \varepsilon(x, t) }$, along with its spatial correlations, is a key physical quantity to characterize the emergence of localized, coherent structures~\cite{Brown_1996, Daumont_1997}. In particular, considering the time average
\begin{equation}
\label{eqn:S8}
\bar{\varepsilon}(x) = \frac{1}{\mathcal{T}} \int_0^\mathcal{T} \varepsilon(x, t) dt ,
\end{equation}
where ${ \mathcal{T} }$ is the observation time, the following spatial correlation function is introduced
\begin{equation}
\label{eqn:S9}
C_{\bar{\varepsilon}} (\mathcal{X}) \propto \frac{\left\langle \int \bar{\varepsilon}(x) \bar{\varepsilon}(x + \mathcal{X}) dx \right\rangle}{{\left\langle \int \bar{\varepsilon}(x) dx \right\rangle}^2} ,
\end{equation}
in which ${ \mathcal{X} }$ is a space displacement, ${ \left\langle ... \right\rangle }$ denotes the ensemble average, and the integrals are performed over the whole spatial domain.

To truly appreciate Eq.~\eqref{eqn:S9}'s significance, it is useful to discuss its behavior in distinct scenarios: (i)~no excitations (i.e., spatially-uniform condition); (ii)~breathers stable both in amplitude and position over long times; (iii)~turbulent kink-antikink regime, with different excitations possibly appearing and wandering/annihilating over time. In the first, trivially-correlated case, the energy is equally distributed among all the system's sites, therefore a flat ${ C_{\bar{\varepsilon}} (\mathcal{X}) }$ is obtained [${ C_{\bar{\varepsilon}} (\mathcal{X}) = 1 }$ under appropriate normalization in Eq.~\eqref{eqn:S9}]. When long-time stable breathers are present, each one results in a prominent ${ \bar{\varepsilon}(x) }$ spike of width ${ \lambda_b = 1 / \sqrt{1 - \omega^2} }$, with ${ \lambda_b }$ being the characteristic length of an unperturbed breather with frequency ${ \omega_b = \omega }$ (${ \omega }$ is the frequency of the sinusoidal force, see above). Thus, ${ C_{\bar{\varepsilon}} (\mathcal{X}) }$ peaks at ${ \mathcal{X} = 0 }$, and it decays to $ 1 $ with the typical scale ${ \lambda_b }$ as ${ \mathcal{X} }$ is increased. In the third situation, a nearly spatially-homogeneous ${ C_{\bar{\varepsilon}} (\mathcal{X}) \approx 1 }$ is restored, but the underlying reason is rather different from that of (i). Many solitonic excitations are indeed observed temporarily in the system, but due to thermal agitation they do not leave long-lasting marks on a few (random) spots. On average, see Eq.~\eqref{eqn:S8}, the spatial domain is roughly explored in a uniform way by these strongly fluctuation-driven transients.

\begin{figure}[t!!]
\includegraphics[width=0.4\columnwidth]{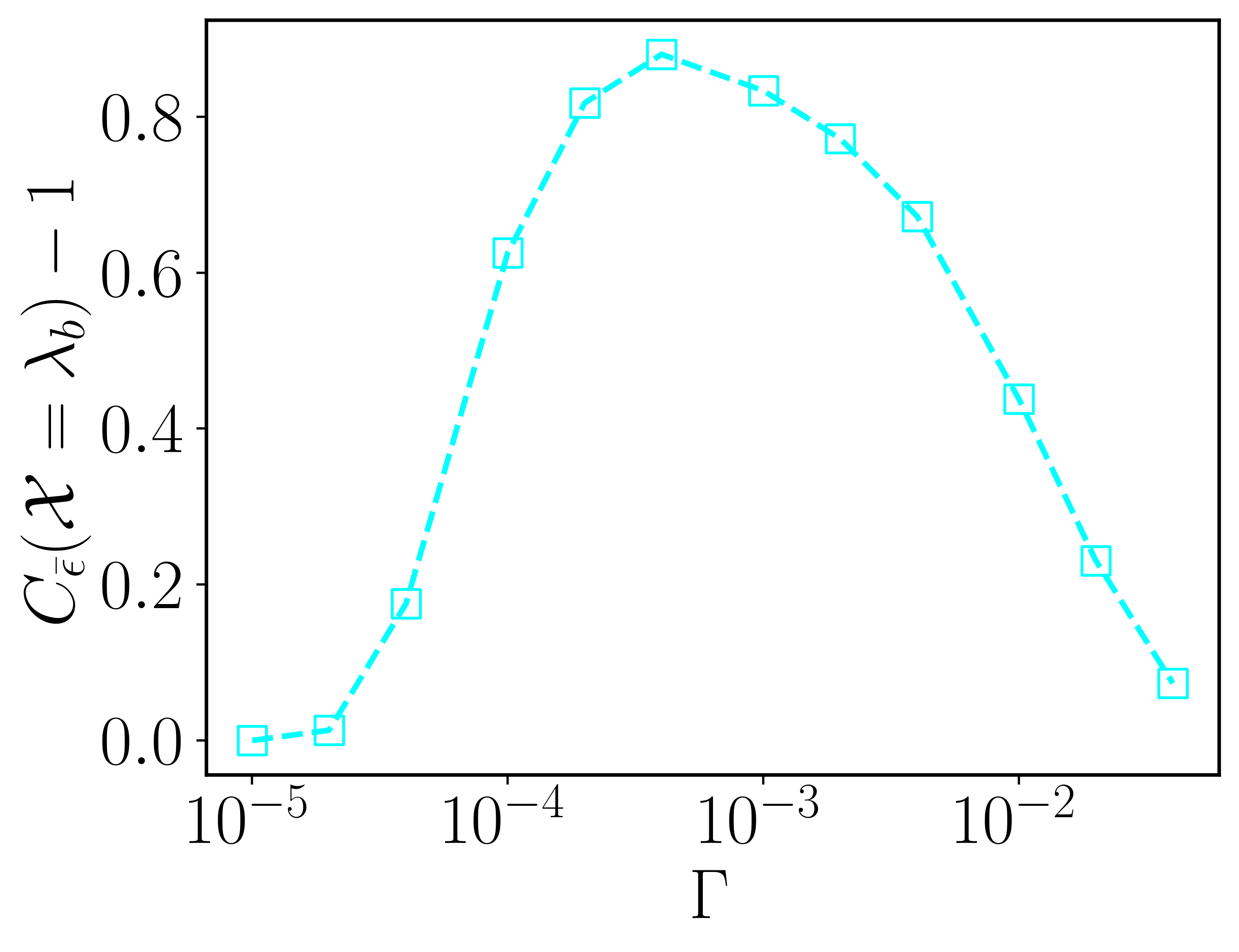}
\caption{Energy-based coefficient of spatial correlation versus ${ \Gamma \in [10^{-5}, 4 \times 10^{-2}] }$. Parameter values: ${ \Delta x = \Delta t = 0.01 }$, ${ l = 50 }$, ${ \mathcal{T} = 500 }$, ${ \alpha = 0.2 }$, ${ \omega = 0.6 }$, ${ \eta = 0.59 }$, and ${ N = 1000 }$.}
\label{fig:S1}
\end{figure}

In light of the above, it is natural to evaluate the system's response in terms of ${ C_{\bar{\varepsilon}} (\mathcal{X} = \lambda_b) }$, as a function of ${ \Gamma }$. As displayed in Fig.~\ref{fig:S1} [i.e., Fig.~2(b) in the main text, reported here for the reader's convenience], such a quantity behaves nonmonotonically versus ${ \Gamma \in [10^{-5}, 4 \times 10^{-2}] }$. The values of the remaining parameters are: ${ \Delta x = \Delta t = 0.01 }$, ${ l = 50 }$ (system length), ${ \mathcal{T} = 500 }$, ${ \alpha = 0.2 }$, ${ \omega = 0.6 }$, ${ \eta = 0.59 }$, and ${ N = 1000 }$ (number of realizations). At each integration step, the phase ${ \varphi }$ is regularized via moving averages before calculating ${ \varphi_t }$ and ${ \varphi_x }$ needed to evaluate ${ \varepsilon (x, t) = (\varphi_t^2 + \varphi_x^2)/2 + 1 - \cos \varphi }$. Clearly, the previously mentioned cases (i), (ii), and (iii) correspond to the phenomenology observed at low, intermediate, and high noise strengths, respectively. Therefore, the trend in Fig.~\ref{fig:S2} is well-understood on physical grounds.

In short, an appropriate amount of environmental noise can enhance the system's sensitivity to the external force, leading to nontrivial spatial correlations.

\section{Typical timescale of the solitonic formation events}

\begin{figure}[b!!]
\includegraphics[width=0.4\columnwidth]{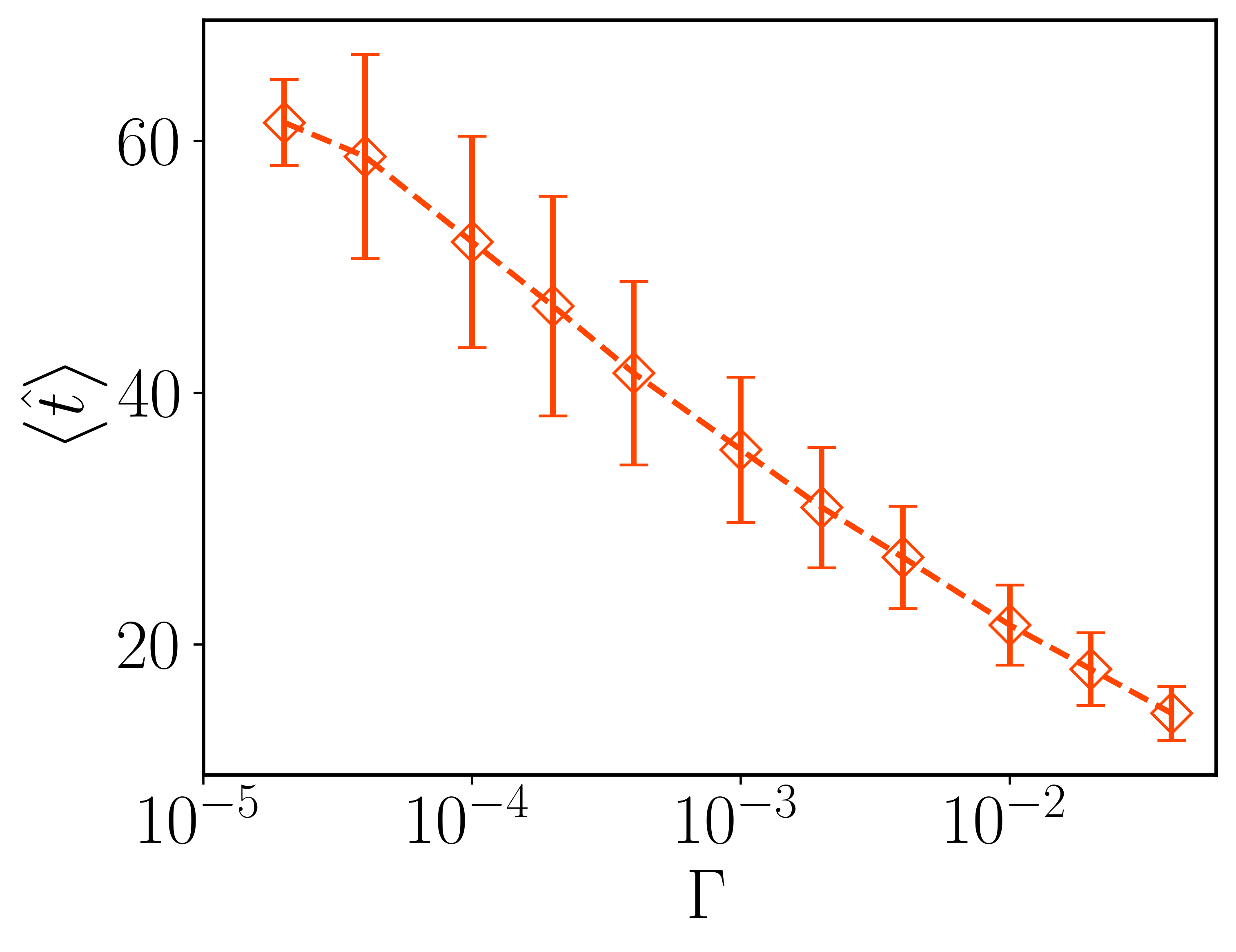}
\caption{Average time ${ \left\langle \hat{t} \right\rangle }$ and the corresponding standard deviation as a function of the noise amplitude ${ \Gamma \in [10^{-5}, 4 \times 10^{-2}] }$. Parameter values: ${ \Delta x = \Delta t = 0.01 }$, ${ l = 50 }$, ${ \mathcal{T} = 500 }$, ${ \alpha = 0.2 }$, ${ \omega = 0.6 }$, ${ \eta = 0.59 }$, and ${ N = 1000 }$.}
\label{fig:S2}
\end{figure}
The noise strength is reasonably expected to influence the typical timescale of the random emergence of the solitonic states (breathers and kink-antikink couples). To examine this, focusing on the fraction of realizations where at least one generation event takes place, one can record the time ${ \hat{t} }$ at which the first excitation with amplitude greater or equal to ${ A_b = 4 \arctan \left( \sqrt{1 - \omega^2} / \omega \right) }$ is observed, with ${ A_b }$ being the amplitude of an unperturbed breather at frequency ${ \omega_b = \omega }$.

Figure~\ref{fig:S2} shows the average time ${ \left\langle \hat{t} \right\rangle }$ and the corresponding standard deviation as a function of ${ \Gamma \in [10^{-5}, 4 \times 10^{-2}] }$. The values of the remaining parameters are ${ \Delta x = \Delta t = 0.01 }$, ${ l = 50 }$, ${ \mathcal{T} = 500 }$, ${ \alpha = 0.2 }$, ${ \omega = 0.6 }$, ${ \eta = 0.59 }$, and ${ N = 1000 }$. A clear result is found: for higher noise amplitudes, on average, nonlinear modes appear in earlier stages of the simulations.